\documentclass{article} 
\usepackage{iclr2019_conference,times}
\iclrfinaltrue

\usepackage{amsmath,amsfonts,bm}

\def\eqref#1{equation~\ref{#1}}

\def\1{\bm{1}}

\DeclareMathAlphabet{\mathsfit}{\encodingdefault}{\sfdefault}{m}{sl}
\SetMathAlphabet{\mathsfit}{bold}{\encodingdefault}{\sfdefault}{bx}{n}

\usepackage{hyperref}
\usepackage{enumitem}
\usepackage{listings}
\usepackage{booktabs}
\usepackage{url}
\usepackage{graphicx}
\usepackage[font=small,labelfont=bf]{caption}

\usepackage{authblk,etoolbox}
\DeclareMathAlphabet{\pazocal}{OMS}{zplm}{m}{n}
\newcommand{\symT}{\pazocal{T}}

\makeatletter
\patchcmd{\@maketitle}{center}{flushleft}{}{}
\patchcmd{\@maketitle}{center}{flushleft}{}{}
\patchcmd{\@maketitle}{}{}{}{}

\def\maketitle{{%
  \renewenvironment{tabular}[2][]
    {\begin{flushleft}}
    {\end{flushleft}}
  \AB@maketitle}}
\makeatother

\title{A Knowledge Graph-based Approach for \\Exploring the U.S. Opioid Epidemic}

\author[1]{\textbf{Maulik R. Kamdar}}
\author[1]{\textbf{Tymor Hamamsy}}
\author[2]{\textbf{Shea Shelton}}
\author[2,3]{\textbf{Ayin Vala}}
\author[2]{\textbf{Tome Eftimov}} 
\author[4]{\\\textbf{James Zou}}
\author[2,4]{\textbf{Suzanne Tamang}}

\affil[1]{Biomedical Informatics Training Program, Stanford University, USA}
\affil[2]{Center for Population Health Sciences, Stanford University, USA}
\affil[3]{Foundation for Precision Medicine, San Francisco, USA}
\affil[4]{Department of Biomedical Data Science, Stanford University, USA}
\affil[ ]{\texttt{\{maulikrk,tymor,sheakat,ayinv,teftimov,jamesz,stamang\}@stanford.edu}}

\begin{document}

\maketitle

\vspace{-4mm}
\begin{abstract}
The United States is in the midst of an opioid epidemic with recent estimates indicating that more than 130 people die every day due to drug overdose. The over-prescription and addiction to opioid painkillers, heroin, and synthetic opioids, has led to a public health crisis and created a huge social and economic burden. Statistical learning methods that use data from multiple clinical centers across the US to detect opioid over-prescribing trends and predict possible opioid misuse are required. However, the semantic heterogeneity in the representation of clinical data across different centers makes the development and evaluation of such methods difficult and non-trivial. We create the Opioid Drug Knowledge Graph (ODKG) -- a network of opioid-related drugs, active ingredients, formulations, combinations, and brand names. We use the ODKG to normalize drug strings in a clinical data warehouse consisting of patient data from over 400 healthcare facilities in 42 different states. We showcase the use of ODKG to generate summary statistics of opioid prescription trends across US regions. These methods and resources can aid the development of advanced and scalable models to monitor the opioid epidemic and to detect illicit opioid misuse behavior.  Our work is relevant to policymakers and pain researchers who wish to systematically assess factors that contribute to opioid over-prescribing and iatrogenic opioid addiction in the US.
\end{abstract}

\vspace{-1mm}
\section{The Opioid Epidemic in the United States}
\vspace{-1mm}
The opioid abuse epidemic is one of the most challenging public health challenges that our nation has ever faced. The US FDA declared the over-prescribing of opioid painkillers to be a leading cause of the astronomical rise in opioid addiction, with 64,000 overdose deaths in 2016 and 2 million people currently addicted ~\citep{instance1290}. Under current conditions, the annual number of opioid overdose deaths in US is projected to reach nearly 82,000 by 2025, resulting in approximately 700,000 deaths from 2016 to 2025 ~\citep{chen2019prevention}.  With eight states reporting more opioid prescriptions than residents ~\citep{Reuben2015}, modifying current prescribing practices is an important strategy for reducing surging rates of opioid overdoses in the US.  

To gain a better understanding on opioid prescription trends across the entire United States and to predict possible opioid abuse behavior in patients, it is imperative to build statistical models using clinical data from multiple healthcare sites in the US. However, due to the vast semantic heterogeneity that still exists between different clinical systems ~\citep{sujansky2001heterogeneous}, and the lack of use of standard terminologies to encode clinical features (e.g., patient medications), developing such statistical models is difficult. Raw patient data is often extracted from legacy databases across multiple clinical centers and transformed under a uniform representation format (e.g., Fast Healthcare Interoperability Resources format ~\citep{shickel2018deep} and OMOP Common Data Model ~\citep{hripcsak2015observational}) for use in these machine learning models. This is a burden on the side of the clinical centers and leads to the creation of multiple copies of private and secure patient data. 

Knowledge graphs can aid in the task of normalization of similar entities encoded using different identifiers and enable integration of data from multiple heterogeneous sources. Knowledge graphs are large directed networks of real-world entities and relations between those entities, with a fixed set of semantic classes and properties ~\citep{ehrlinger2016towards}. Knowledge graphs constructed from multiple, heterogeneous pharmacological data sources have been used to predict adverse side effects that manifest on the account of polypharmacy ~\citep{zitnik2018modeling,kamdar2017phlegra}. 

In this paper, we describe our efforts to create an Opioid Drug Knowledge Graph (ODKG). We use the ODKG to normalize drug strings from a data warehouse consisting of electronic medical record (EMR) data that was collected from one vendor with installations in 42 states in the US. We showcase how the ODKG can aid in generating summary statistics on the prescription of different opioids in the US. Finally, we will discuss potential applications of using the ODKG to develop a web-based US compendium that allows for exploring and visualizing opioid prescribing across the US.  Although such tools exist to examine international opioid consumption trends ~\citep{usOpioidStlas,acOpioidAtlas}, there is no such resource for the greater US.

\vspace{-2mm}
\begin{table}[t]
\small
\centering
\setlength{\tabcolsep}{3pt}
\caption{First degree hops from the generic ingredient Morphine in the ODKG, listing few examples of different drug formulations, combinations, and tradenames.}
\label{tab:propnav}
\begin{tabular}{l | p{11.5cm}}
\toprule
\textbf{Property Type} & \textbf{Example Classes}   \\
\toprule
\textit{Part Of} & Atropine / Morphine, Cyclizine / Morphine, Morphine / Naltrexone \\
\hline
\textit{Has Tradename} & MS Contin, EMbeda, Avinza, Duramorph, Kadian \\
\hline
\textit{Has Form} & Morphine Hydrochloride, Morphine Sulphate, Morphine Tartrate \\ 
\hline
\textit{Ingredient Of} & Morphine Injectable Solution, Morphine Prefilled Syringe, Morphine / Naltrexone Extended Release Oral Tablet, Morphine Sulfate 20 MG/ML, Morphine hydrochloride 40 MG\\
\hline
\bottomrule
\end{tabular}
\vspace{-2mm}
\end{table}

\section{Methods}

\vspace{-1mm}
\subsection{Generation of the Opioid Drug Knowledge Graph}
\vspace{-2mm}
\label{meth:odkg}
We use two terminologies to generate the Opioid Drug Knowledge Graph (ODKG): $\symT_1$) \textbf{ATC} ~\citep{world2000anatomical}: Active ingredients of drugs classified according to their anatomical, therapeutic, and chemical properties, and $\symT_2$) \textbf{RxNorm} ~\citep{liu2005rxnorm}: Standard names for clinical drugs and dosage forms, as well as relations between clinical drugs to their active ingredients, drug components, and related brand names. Both these terminologies are members of the Unified Medical Language System (UMLS) ~\citep{bodenreider2004unified} and are retrieved from the BioPortal repository of biomedical ontologies and terminologies ~\citep{whetzel2011bioportal}. UMLS uses the notion of a Concept Unique Identifier (CUI) to map classes with similar meaning in different terminologies.

In the first step, we use hierarchical reasoning to retrieve all the descendants of five base opioid-related classes in the ATC terminology: (\texttt{N02A}) Opioid analgesics, (\texttt{N01AH}) Opioid anesthetics, (\texttt{R05DA}) Opium alkaloids and derivatives, (\texttt{N07BC}) Drugs used in opioid dependence, and (\texttt{A06AH}) Peripheral opioid receptor antagonists. That is, we retrieve active ingredients of opioid drugs. Using the UMLS CUI mappings we retrieve classes related to these opioid drug ingredients from the RxNorm terminology. RxNorm terminology has different classes pertaining to drug formulations (e.g., Morphine Sulfate 50 Mg), drug combinations (e.g., Atropine/Morphine), trade names, etc. Each RxNorm class may have a distinct CUI code and an RxCUI code. We retrieve these class relations through a fixed set of properties: \textit{Ingredients Of}, \textit{Has Form}, \textit{Form Of}, \textit{Part Of}, \textit{Ingredient Of}, \textit{Consists Of}, \textit{Constitutes}, \textit{Has Tradename}, and \textit{Precise Ingredient Of}. 

\vspace{-1mm}
\subsection{EMR Data Collection}
\vspace{-2mm}

The Electronic Medical Record (EMR) data is de-identified and is aggregated in structured form from more than 400 hospitals and healthcare facilities from across 42 states in US (59\% South, 17\% West, 13\% Midwest, 12\% Northeast) during 2009-2016. The mix of hospitals consist of large and small facilities in both urban (87\%) and rural (13\%) locations. Roughly 98\% of providers submit both inpatient and outpatient data. The dataset includes hundreds of variables, such as each patient's demographics, diagnoses, procedures and prescribed medications; also, the type and location of facilities and utilization costs. Data from psychiatric admission facilities were excluded from the dataset due to HIPAA rules ~\citep{annas2003hipaa}. The aggregated EMR data is stored in the Google BigQuery Analytics data warehouse ~\citep{tigani2014google}.

\vspace{-1mm}
\subsection{Normalization of Medication Information in the EMR Data Warehouse}
\vspace{-2mm}
Drugs administered to patients in different healthcare sites across the US are recorded and stored using site-specific identifiers with drug strings (e.g., \textit{Duramorph 10mg/10ml EA}, \textit{Morphine Sulfate (Concentrate) 10 mg/0.5ml OR Soln}, \textit{Roxanol Liquid 120ml} all include the active ingredient Morphine). We extract a list of 425,059 unique drug strings from the aggregated EMR data warehouse. MedEx is a natural language processing system to extract medication information from clinical free text ~\citep{xu2010medex}. We use MedEx to parse these drug strings and extract drug name (e.g., Morphine Sulfate), strength (e.g., 10 mg/0.5ml), dosage forms (e.g., OR Soln), etc.  Drug names are mapped with corresponding CUIs and RxCUIs and ODKG classes are instantiated with these drug strings. It should be noted that one drug string can be mapped to multiple CUIs or RxCUIs.

\vspace{-1mm}
\section{Results}
\vspace{-1mm}

High resolution visualizations and detailed results for this research are made available online at \url{https://github.com/maulikkamdar/ODKG}.

\vspace{-1mm}
\subsection{Characteristics of the Opioid Drug Knowledge Graph}
\vspace{-2mm}

\begin{figure}[t]
\vspace{-4mm}
\begin{center}
\includegraphics[scale=0.2]{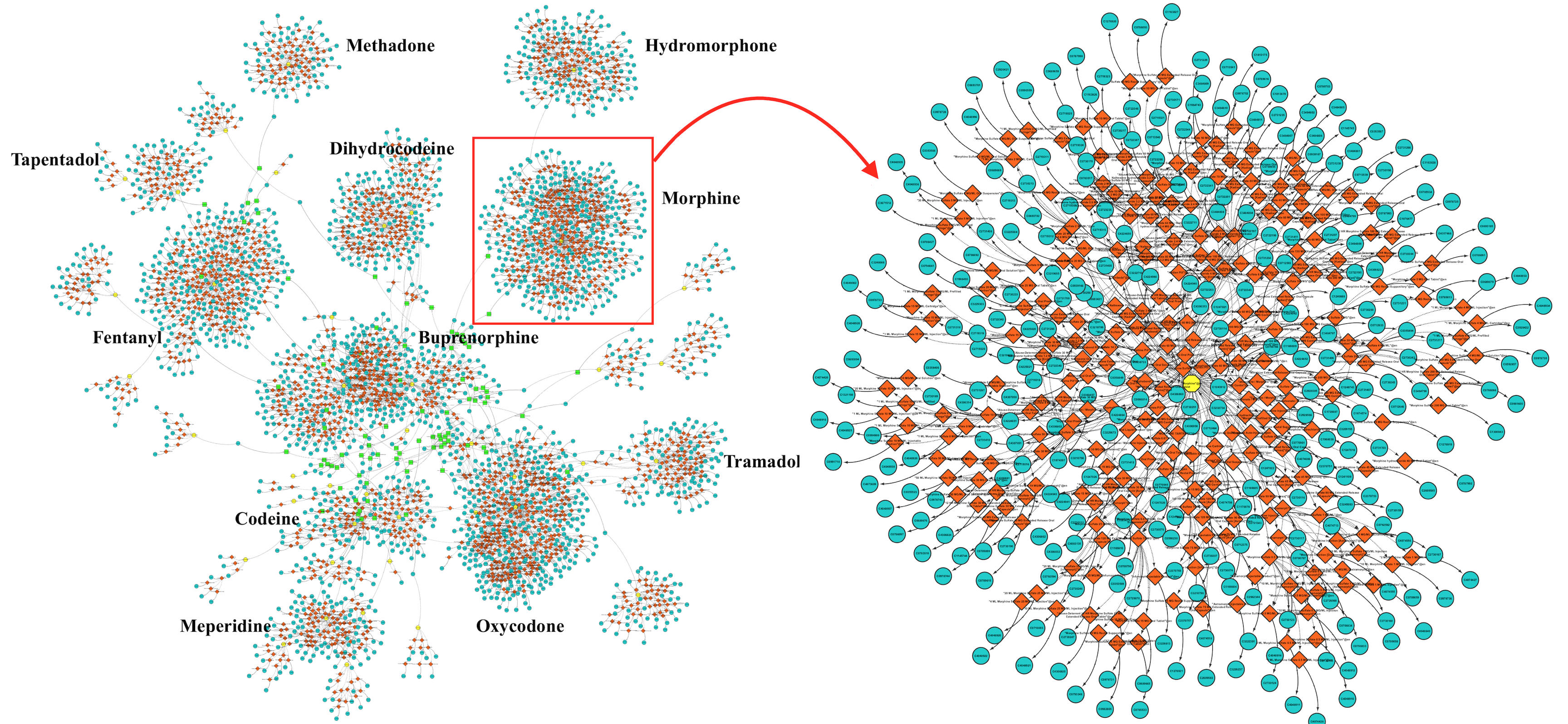}
\end{center}
\caption{\textbf{Opioid Drug Knowledge Graph:} All the classes and their associated relations in the ODKG are visualized using a force-directed network layout. The different classes --- RxNorm drugs, RxNorm generic ingredients, CUIs, and ATC drug classes, are shown as red diamond, yellow square, blue circle, and green square nodes respectively. The directed edges indicate a relation (e.g., \textit{subClassOf}, \textit{hasCUI}, or any property listed in \textbf{Section \ref{meth:odkg}}) between the linked classes. The Morphine-specific network is shown in greater detail. }
\label{fig:opioidkg}
\vspace{-5mm}
\end{figure}

The Opioid Drug Knowledge Graph (ODKG), extracted from the ATC and RxNorm terminologies, has a hierarchical classification backbone with 97 ATC drug classes, 48 generic RxNorm drug ingredients, 4,960 other RxNorm classes (i.e., combinations, formulations, and ingredients), 5,051 CUI nodes and 5,188 RxCUI annotations, and 13,581 semantic relations of the type \textit{subClassOf}, \textit{hasCUI}, or any of the above-mentioned property types (see \textbf{Section \ref{meth:odkg}}). The ODKG is visualized in \textbf{Figure \ref{fig:opioidkg}}, with the Morphine-related community highlighted in more detail. The different classes --- RxNorm drugs, RxNorm generic ingredients, CUIs and ATC drug classes are shown as red diamond, yellow square, blue circle, and green square nodes respectively. A directed edge between two ATC drug classes represents the \textit{subClassOf} relation, whereas a directed edge between an active ingredient and an RxNorm node indicates a relation between those two classes subscribed under a property type. Each RxNorm node is associated with CUI node(s). Representation of clinical knowledge in graphical format enables ease of querying and abstraction of different drug strings. For example, as shown in \textbf{Table \ref{tab:propnav}}, first degree hops from the generic ingredient Morphine enables the retrieval of different drug combinations, tradenames, formulations, and dosages. 

\vspace{-1mm}
\subsection{Efficacy of MedEx for Normalizing Medication Information}
\vspace{-2mm}

There were 425,059 unique drug strings in EMR data warehouse. After extracting medication information using MedEx, 288,983 drug strings are mapped to at least 1 CUI (68\% coverage), and 374,208 drug strings are mapped to at least 1 RxCUI (88\% coverage). The opioid-related drug classes are instantiated with the normalized drug strings. There are 29 opioid-related active ingredients classes in the ODKG which have more than 10 drug strings from the EMR data warehouse instantiated under them. It can be seen in \textbf{Figure \ref{fig:stats}A} that certain opioid painkillers, such as Morphine, Oxycodone, Hydromorphone, as well as synthetic opioids, such as Fentanyl, may have more than 1,000 drug strings instantiated under them. This demonstrates the efficacy of the ODKG toward drug string normalization in clinical data from across multiple centers.

\vspace{-1mm}
\subsection{Summary Statistics across Location and Time}
\vspace{-2mm}
\begin{figure}[t]
\vspace{-4mm}
\begin{center}
\includegraphics[scale=0.33]{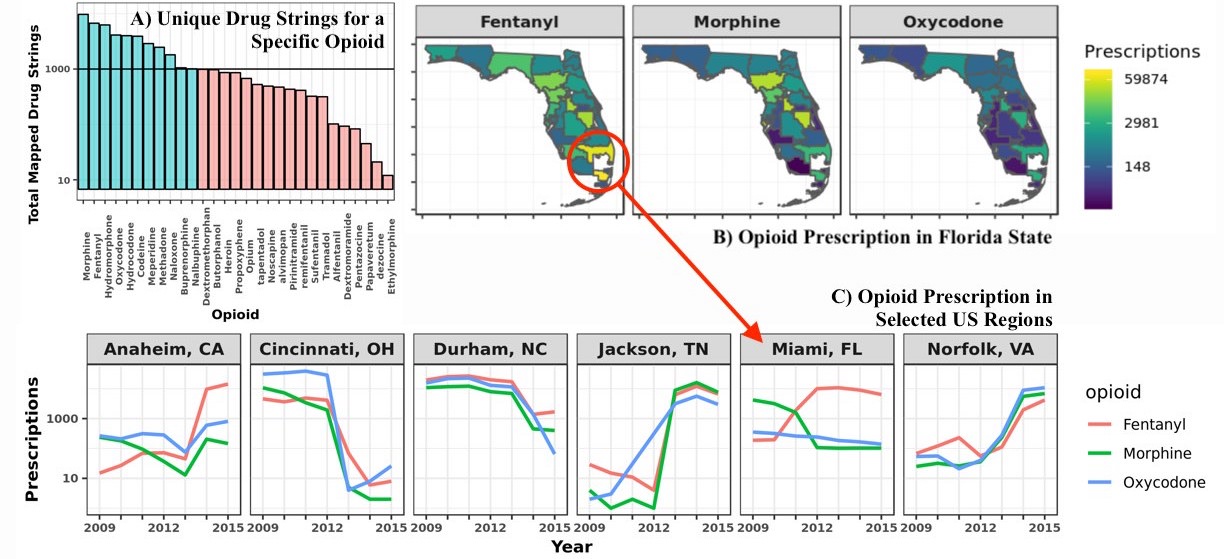}
\end{center}
\caption{\textbf{A)} Total number of unique drug strings in the EMR data warehouse instantiated under the generic opioid classes in the ODKG. \textbf{B)} Florida region-wise visualization of unique prescription occurrences for three types of opioids: Fentanyl, Morphine, and Oxycodone. \textbf{C)} Increase or decrease in opioid prescriptions across selected US regions with more than 10,000 unique prescriptions for a particular opioid at a specific time period.}
\label{fig:stats}
\vspace{-5mm}
\end{figure}

We show a small application of using ODKG in conjunction with the aggregated EMR data warehouse to generate summary statistics of prescriptions of different opioids. Opioid prescriptions are categorized according to different US regions as well as different time periods (as determined through admission year of the patient). \textbf{Figure \ref{fig:stats}B} shows a visualization of the Florida state where it can be observed that the Miami region in Florida has $\approx$ 60,000 unique Fentanyl prescriptions. Moreover, as seen in \textbf{Figure \ref{fig:stats}C}, there is a distinct bump in the rate of Fentanyl prescriptions around 2012. \textbf{Figure \ref{fig:stats}C} also shows selected US regions that demonstrate more than 10,000 unique prescriptions for either Fentanyl, Morphine, or Oxycodone, in a specific year. Such visualizations may be used for hypothesis generation (e.g., increase in opioid prescription for a particular region). 

\vspace{-1mm}
\section{Discussion}
\vspace{-2mm}
Heterogeneous drug names and drug composition pose a significant challenge in performing data science and machine learning to study the opioid epidemic. In this work, we address this challenge by developing the Opioid Drug Knowledge Graph (ODKG), the first knowledge graph that captures how opioid drugs relate to each other. This knowledge graph makes it straightforward to translate medications from diverse electronic medical records into a common set of chemical-dosage features, which subsequently enables a large number of prediction and modeling tasks. 

In order to identify the best strategies to reduce opioid over-prescription and misuse, a better understanding of country and regional consumption patterns, pharmaceutical industry influences, and sociopolitical factors that impact consumption, is needed. Our ODKG will be used to develop a web-based tool that can facilitate visualization of historical patterns and can enable comparisons across opioids, time, and US regions. Since our ODKG was data-driven, we hope to further refine it by consulting with a domain expert and tailor it for specific use cases and end users.  

Additionally, we have identified several next steps for further research. We plan to compare our approach against the OMOP-based approach of transformation of clinical data ~\citep{hripcsak2015observational}. Potential areas of application include the development of dynamic phenotyping methods to visually analyze individual pain medication use profiles, to identify potential risk factors for long-term use, and to detect adverse outcomes for incident user of prescription opioids, specific to the surgical setting, which some experts now consider the new `gateway' to drug abuse.



\bibliography{iclr2019_conference}
\bibliographystyle{iclr2019_conference}

\end{document}